\documentclass[preprint,prX]{elsarticle}
\usepackage[bottom =2.50cm, right= 3.00cm, left=3.00cm, top=3.0cm]{geometry}                		
\usepackage{multicol}
\usepackage[parfill]{parskip}    		        
\usepackage{graphicx}				
\usepackage{amssymb}
\usepackage{lscape}
\usepackage{longtable}
\usepackage{enumitem}
\usepackage{lineno}
\usepackage{amsmath}
\usepackage{multirow}
\usepackage{amsmath, amsthm, amssymb}

\begin{document}
\title{Expected Sensitivity to Galactic/Solar Axions\\ and Bosonic Super-WIMPs based on the Axio-electric Effect\\ in Liquid Xenon Dark Matter Detectors}
\author{K.~Arisaka\footnote{Corresponding author: arisaka@physics.ucla.edu}}
\author{P.~Beltrame}
\author{C.~Ghag\footnote{Current address:  Department of Physics and Astronomy, University College London, London WC1E 6BT, UK}}
\author{J.~Kaidi}
\author{K.~Lung}
\author{\\A.~Lyashenko}
\author{R.~D.~Peccei}
\author{P.~Smith}
\author{K.~Ye\footnote{Current address: Yuanpei College, Peking University, China}}

\address{Department of Physics and Astronomy, University of California, Los Angeles,\\475 Portola Plaza, Los Angles, CA 90095, USA}
\date{}
\begin{abstract}

We present systematic case studies to investigate the sensitivity of axion searches by liquid xenon detectors, using the axio-electric effect (analogue of the photoelectric effect) on xenon atoms. Liquid xenon is widely considered to be one of the best target media for detection of WIMPs (Weakly Interacting Massive Particles which may form the galactic dark matter) using nuclear recoils. Since these detectors also provide an extremely low radioactivity environment for electron recoils, very weakly-interacting low-mass particles ($<$ 100 keV/c$^2$), such as the hypothetical axion, could be detected as well $-$ in this case using the axio-electric effect. Future ton-scale liquid Xe detectors will be limited in sensitivity only by irreducible neutrino background (pp-chain solar neutrino and the double beta decay of $^{136}$Xe) in the mass range between 1 and 100 keV/c$^2$. Assuming one ton-year of exposure, galactic axions (as non-relativistic dark matter) could be detected if the axio-electric coupling $g_{Ae}$  is greater than $10^{-14}$ at 1 keV/c$^2$ axion mass (or $10^{-13}$ at 100 keV/c$^2$).  Below a few keV/c$^2$, and independent of the mass, a solar axion search would be sensitive to a coupling $g_{Ae} \sim10^{-12}$.  This limit will set a stringent upper bound on axion mass for the DFSZ and KSVZ models for the mass ranges $m_A < 0.1$ eV/c$^2$ and $< 10$ eV/c$^2$, respectively.  Vector-boson dark matter could also be detected for a coupling constant $\alpha'/ \alpha > 10^{-33}$ (for mass 1 keV/c$^2$) or $> 10^{-27}$ (for mass 100 keV/c$^2$).

\end{abstract}


\maketitle

\section{Introduction}

Early conjectures about the nature of the galactic dark matter, estimated from the stellar velocity distribution to have a density $\sim$0.3 GeV/cm$^3$ at the location of the solar system, focused on two main ideas: (a) that it may consist of WIMPs, Weakly Interacting Massive Particles, such as the lightest supersymmetric particle~\cite{Goodman:1985}, or (b) that it may consist of a very light boson, called the axion, associated with a symmetry proposed by Peccei and Quinn~\cite{Peccei:1977,Weinberg:1978,Wilczek:1978} with a symmetry-breaking energy scale $\sim 10^{12}$ GeV. Extensive studies of the possible role of axions in astrophysics and primordial cosmology have been made by Khlopov et al~\cite{Khlopov:1999,Khlopov:2004,Khlopov:2012}. This particle was originally termed ``invisible'' because of its predicted extremely weak interaction with normal matter. However, in 1983 it was shown by Sikivie~\cite{Sikivie:1983} that these invisible axions could, in principle, be detected via their coupling to two photons (Primakoff effect) by conversion to a microwave photon, with energy equal to the axion mass, in a magnetic field. 

Sikivie's paper also renewed interest in the detection of solar axions. If the symmetry-breaking scale is lower than that needed for dark matter axions, axions produced in the sun might be detectable through their coupling to photons, electrons and nuclei~\cite{Weinberg:1978,Cheng:1987}. These solar axions with energies peaking at around a few keV (independent of axion mass) can be converted by a magnetic field to X-rays and detected in this fashion, providing evidence that axions exist even if they are not the dark matter in the Universe. In this context, another important idea was suggested which was  to use the analogue of the photoelectric effect, termed the axio-electric effect, to provide conventional detection by ionization of atoms~\cite{Dimopoulos:1986}. This principle was immediately applied by Avignone et al.~\cite{Avignone:1987} to their existing underground germanium detector, setting dimensionless coupling limits $\sim10^{-9}$ from the statistical absence of a signal in their detector background.  Further detection ideas for both solar axions and dark matter axions, as well as Axion-Like Particles (ALPs) in which the connection between the mass and the symmetry breaking scale is relaxed, are reviewed by Smith $\&$ Lewin~\cite{SmithLewin:1990} and by Rosenberg and van Bibber~\cite{Rosenberg:2000,Rosenberg:2010}.

Subsequently, experiments searching for WIMPs and axions diverged and specialized: axion searches concentrated on improved cavity microwave experiments for dark matter such as the Axion Dark Matter experiment (ADMX)~\cite{Rosenberg:2000,Rosenberg:2010} and solar axion telescopes such as the CERN Axion Solar Telescope (CAST)~\cite{Andriomonje:2007} for solar axions, while WIMP searches have concentrated on nuclear recoil detection using ionization, scintillation, bolometric methods, and combinations of these~\cite{Gaitskell:2004}, using targets of increasing mass and increasingly low background~\cite{Arisaka:2012,Arisaka:2009}. These improvements in background, arising both from reduction in detector radioactivity levels and from the use of outer regions of the target as self shielding~\cite{Arisaka:2012}, have recently increased interest in using the same WIMP detectors also to detect, or set new limits on, both dark matter and solar axions, in addition to other new types of postulated particle which may be the galactic dark matter.~\cite{Pospelov:2008}

Avignone~\cite{Avignone:2009} has recently discussed the use of future large scale Ge detector arrays, using the axio-electric effect, to improve the axion-photon coupling limit to the $10^{-10}$ level, and Avignone et al.~\cite{AvignoneCreswick:2009} have discussed the use of annual modulation of the axio-electric effect in future large scintillator arrays to probe possible axion-photon oscillation phenomena. The cryogenic CDMS dark matter experiment (Ahmed et al.~\cite{Ahmed:2009}) has set axion coupling limits via Primakoff conversion and coherent Bragg diffraction in Germanium crystal ionization detectors. Derbin et al.~\cite{Derbin:2011,Derbin:2012} have used a Si(Li) crystal to detect axio-electric signals from both Si and Tm target crystals, and in turn to set limits on the solar axion flux, thus providing limits on both axion-electron coupling and axion mass. Suzuki et al.~\cite{Suzuki:2012} are exploring limits on galactic/solar axions as well as WIMP dark matter from the single phase liquid Xe XMASS experiment. Finally, Pospelov et al.~\cite{Pospelov:2008} showed that variants of photoelectric or axio-electric absorption of dark matter can provide sensitivity to superweak couplings from other more general bosonic dark matter candidates.

In this paper we investigate the improvements in detection sensitivity which would be obtainable with both existing and future large-scale (0.1-10 ton) ultra-low background liquid Xe targets, especially the 2-phase liquid/gas instruments, showing the sensitivity and confidence limits for axion and other boson couplings as a function of background level, energy resolution, and target exposure. Detection of axions by the axio-electric effect measures specifically the coupling of axions to electrons, and does not depend on the existence of an axion-photon coupling. Liquid xenon detectors can perform searches for both types of axion particle - dark matter (galactic) axions and solar axions. For dark matter axions, which are non-relativistic, the mass of the axion would be fully converted to the electron recoil energy, giving a signal that is monochromatic (with the exception of detector energy resolution effects). Solar axions are produced with a spectrum peaking at around a few keV, regardless of their rest mass, so that detection of solar axions is possible even though their intrinsic mass may be extremely small. We consider the basic theory of signal from galactic axions in Section 2, and solar axions in Section 3.  This is followed in Section 4 with case studies examining achievable sensitivity to axion interactions with existing and future liquified Xe detectors.  In Section 5 we explore the potential of simultaneous searches for further types of dark matter, in particular, keV-scale vector boson dark matter.

\section{Galactic Axion Search by Axio-electric Effect}
\subsection{Relationship of axio-electric effect to photoelectric effect}

Derevianko et al.~\cite{Derevianko:2010} and Derbin et al.~\cite{Derbin:2012} showed that the cross section $\sigma_A (E)$ for the axio-electric effect is proportional to the photoelectric cross section, $\sigma_{pe}(E)$, and is given by the formula:
\begin{equation}
\label{eqn:xs_A}
\sigma_A(E)=\sigma_{pe}(E)\frac{{g_{Ae}}^2}{\beta}{\frac{3{E}^2}{16\pi\alpha{m_e}^2}}(1-\frac{\beta}{3})
\end{equation}
where $E$ is the axion total energy, $\beta$ is the axion velocity divided by the velocity of light, and $g_{Ae}{=}2m_{e}/f_{a}$ is the dimensionless axion-electron coupling constant with the strength of the standard model-axion interaction $f_{a}$.

Under the assumption that axions constitute the whole of the galactic dark matter density $\rho_{DM}\sim 0.3$ ${\rm{GeV/cm^{3}}}$, then the total flux of dark matter axions $\Phi{=}\rho_{DM}v_{a}/m_{A}$ is given by 

\begin{equation}
\label{eqn:phi}
\Phi[\rm{/cm^{2}/s}]=\left(\frac{9\times10^{15}}{ \mathit{m}_A }\right)\beta_{m}
\end{equation}
where $m_A$ is the axion mass in $\rm{keV/c^{2}}$ and $\beta_{m}$ is the mean velocity of the axion distribution relative to the Earth $\beta_{m}\sim 10^{-3}$.
The interaction rate $R[\rm{/kg/day}]$ is given by
\begin{equation}
\label{eqn:rate_DM}
R[{\rm{/kg/day}}]=\Phi\times 86400 \times\frac{6\times 10^{23}}{A}\sigma_A(E).
\end{equation}

From the preceding equations, the rate can be expressed as~\cite{Pospelov:2008}

\begin{equation}
\label{eqn:rate_DM_1}
R [\mathrm{/kg/day}]=\left(\frac{1.29\times10^{19}}{A}\right){{g_{Ae}}^2}m_A[\rm{keV/c^{2}}]\sigma_{pe}[\rm{barns}]
\end{equation}
 where the atomic number averages $A$ = 131 for natural xenon. It should be noted that in the non-relativistic case the rate does not depend on the galactic axion velocity, since for each element of flux with velocity $\beta$, there is a cancellation with the 1/$\beta$ of Eq. (1). Thus uncertainties in the axion velocity distribution do not affect the result.

A positive axion signal would be seen as a peak at the axion mass, spread by the energy resolution.  In the absence of peaks, statistical limits can be placed on the number of axion events in the experimental background for a given experimental signal energy span, from which a limit can be placed on $\sigma_A$ and hence on ${g_{Ae}}^2$.

\begin{figure}[htp]
\centering
\includegraphics[width=14cm] {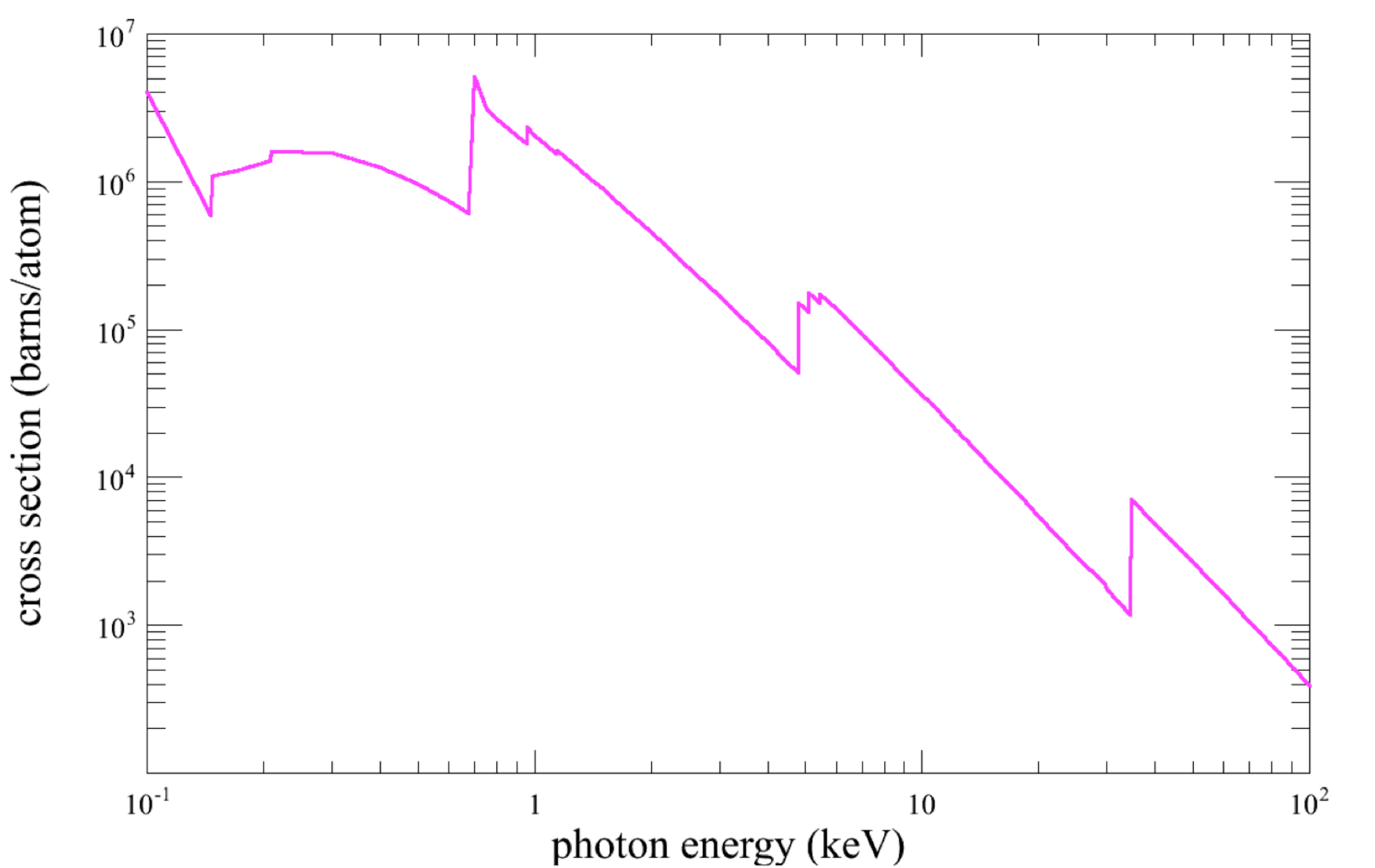}
\caption{Photo-electric cross section on Xenon atom~\cite{Veigele:1973}.}\label{fig:PhotoelectricCrossSection}
\end{figure}

\begin{figure}[htp]
\centering
\includegraphics[width=14cm] {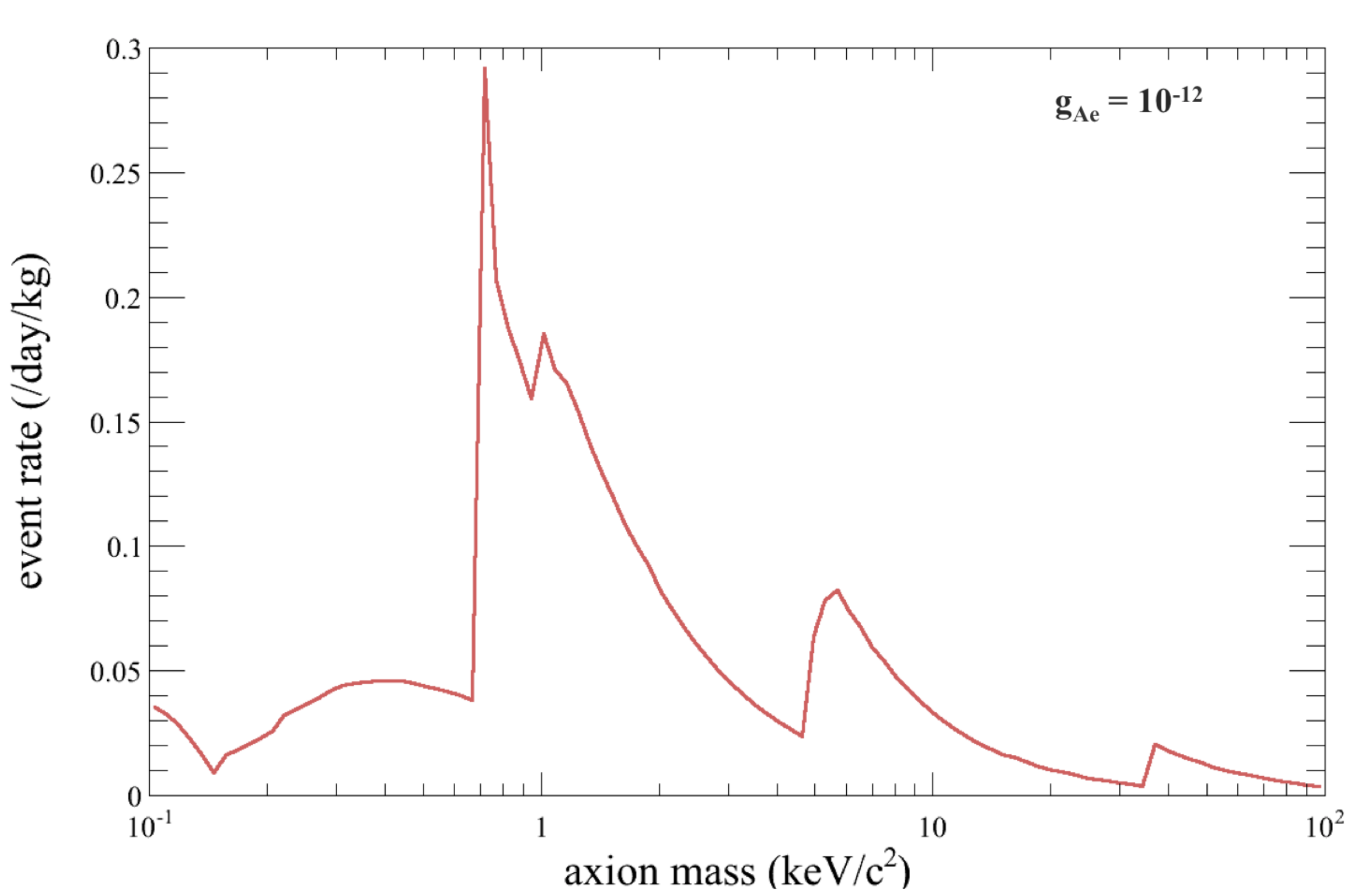}
\caption{Event rate [/day/kg] of DM (galactic) axion as a function of axion mass, corresponding to an axio-electric coupling $g_{Ae}{=} 1\times10^{-12}$.}\label{fig:DMEventRate}
\end{figure}

The photoelectric cross section for xenon, used to obtain the axio-electric cross section from Eq.~\ref{eqn:xs_A}, is shown in Figure~\ref{fig:PhotoelectricCrossSection}~\cite{Veigele:1973}.  Figure~\ref{fig:DMEventRate} shows the corresponding event rate (/kg/day) as a function of axion mass.  In practice, the dark matter axion mass range is limited at low masses (at $\sim$ keV/c$^2$) by the experimental energy threshold, and at high masses by the increasing background beyond $\sim$100 keV/c$^2$. Note that conventional axions, assumed to be the dark matter of the Universe, have an expected mass range in the micro-eV range and so will not be probed by this technique. However the axion-like particles discussed by Pospelov et al ~\cite{Pospelov:2008} can have masses in the keV range and be the dark matter in the Universe. It is these latter axion variants that WIMP detectors can discover or provide strong constraints for their existence.

\subsection{Procedure for experimental limits in xenon detectors}
\label{sec:ExperimentalLimits}
For analysis of the experimental background, we introduce the quantities contained in Table~\ref{table:nonlin}. Note that background rates are given in units of ``dru'' (1 dru = 1 event/kg/day/keV) for convenience, as is the convention for WIMP dark matter search experiments.

  \begin{table}
  \caption{List of quantities adopted and used throughout this work}
  \centering
\begin{tabular}{l c c}
\hline\hline
Quantity&symbol&unit\\\hline
  Mass of axion & $m_{A}$ & [keV/c$^{2}$] \\
  Axio-electric coupling & $g_{Ae}$ &  \\
  Photoelectric cross section & $\sigma_{pe}$ & [barn/atom] \\
  Axio-electric cross section & $\sigma_{A}$ & [barn/atom] \\
  Data-taking live-time & $T$ & [days] \\
  Fiducial mass of xenon & $W$ & [kg] \\ 
  Energy of recoil electron & $E$ & [keV] \\ 
  Energy resolution at 1 keV & $b$ & $[\sqrt{\mathrm{keV}}]$ \\
  90\% C.L. upper limit of signal events in selected energy span & $S$ & [events] \\
  Differential background event rate & $dN/dE$ & [/kg/day/keV (dru)] \\ 
  Background events in selected energy span & $N$ & [events] \\ 
  Significance level (signal to noise ratio) & $r_{sn}$ & [standard deviations] \\
  \hline
\end{tabular}
\label{table:nonlin}
\end{table}

\normalsize Considering a Poisson distribution of the detected signals, the energy resolution $\sigma_E$ at an energy $E$ is given by
\begin{equation}
\label{eqn:resolution}
\frac{\sigma_E}{E}=\frac{b}{\sqrt{E}}
\end{equation}
where $b$ is the numerical value of the resolution at $E$ = 1 keV. In the case of a single phase detector (such as XMASS)~\cite{Suzuki:2012}, $b$ is given by the Poisson distribution of scintillating photons (S1), whereas in case of the dual phase TPC detector (such as XENON100 and LUX), $b$ is given either by the secondary ionization signal of drifted electrons (S2) or by the combined energy scale of the S1 and the S2~\cite{Arisaka:2012_2D}. 
 
We first consider a $\pm{2 \sigma}$ window around the monochromatic peak of a given electron recoil energy. The signal region is defined by this window unless the peak energy minus 2$\sigma$ was less than the detector energy threshold, in which case the lower energy bound of the region is set at the energy threshold.  For a target exposure of $T\times{W}$, the total number of background events in the signal region is 

\begin{equation}
\label{eqn:nBG}
N=4b\sqrt{E}\times\frac{dN}{dE}\times{TW}.
\end{equation}

Assuming Poisson fluctuation of the background $N$, we can define the signal-to-noise ratio of the signal as the quantity
\begin{equation}
\label{eqn:significance}
r_{sn}=\frac{S}{\sqrt{N}}
\end{equation}

where $S$ is given by the Feldman-Cousins 90\% upper confidence limit of signal events for the background only case ($n_{obs}{=}n_{exp}{=}N$)~\cite{Feldman:1998} such that $r_{sn}{=}\sim1.7$ for $N>50$. Thus by equating $S$ from Eq.~\ref{eqn:significance} to $R\times{T}\times{W}$ from Eq.~\ref{eqn:rate_DM_1}, we can evaluate the $90\%$ confidence limit on $g_{Ae}$ in Eq.~\ref{eqn:xs_A}.

To see the relative importance of the various experimental parameters, it is useful to combine the above equations to obtain explicitly an expression for $g_{Ae}$ in terms of those parameters and the confidence limit parameter $r_{sn}$,

\begin{equation}
\label{eqn:DMUpperLimit}
g_{Ae}(90\% CL)=\frac{5.6\times10^{-9}}{{m_A}^{3/8}}\left(\frac{r_{sn}}{\sigma_{pe}}\right)^{1/2}\left(\frac{b\times(dN/dE)}{TW}\right)^{1/4}
\end{equation}

showing that improvements in background, target mass, live time and resolution improve limits on $g_{Ae}$ only as the fourth root of those experimental quantities. 

Examples of the application of this to specific detectors with assumed operational parameters are shown in Section~\ref{sec:casestudies}.

\section{Solar Axion Search by Axio-electric Effect}
\subsection{Flux of solar axions}

Derbin et al.~\cite{Derbin:2011,Derbin:2012} have calculated the flux of axions from the sun arising from the two principal processes \begin{math}
\gamma + e^-{\rightarrow}e^-+A\end{math} (Compton) and \begin{math}e^-+Z{\rightarrow}e^-+A\end{math} (Bremsstrahlung), the nuclei considered for Z being $^1$H, $^4$He, $^3$He, $^{12}$C, $^{14}$N and $^{16}$O. We consider only the dominant two processes, shown in Figure 3, neglecting smaller production processes like the Primakov effect and nuclear de-excitation.

\begin{figure}[htp]
\centering
\includegraphics[width=12cm] {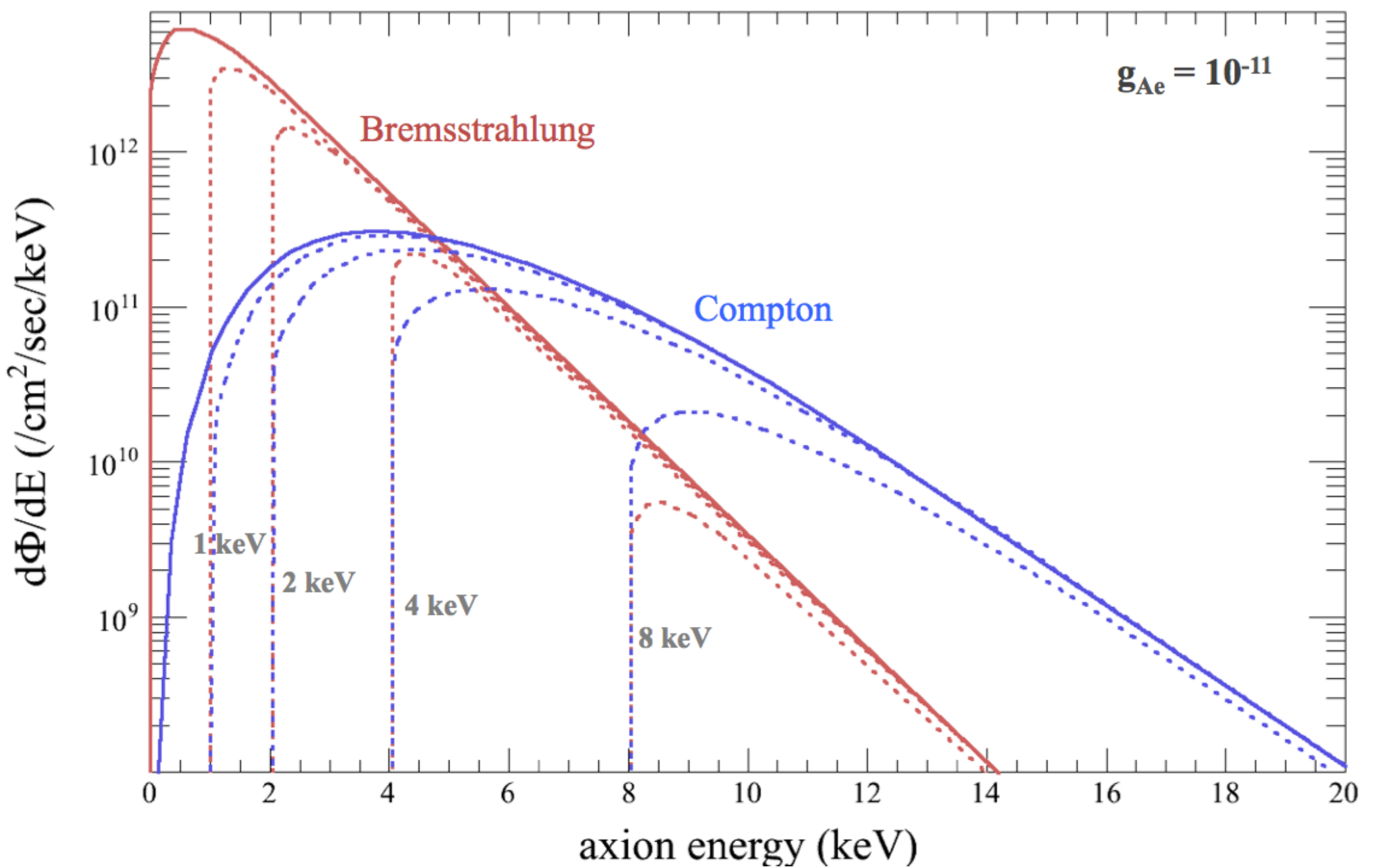}
\caption{Differential energy spectrum solar axion flux (from Derbin et al.~\cite{Derbin:2012})  showing separately axion flux from Bremsstrahlung (red) and Compton scattering (blue). The absolute flux corresponds to an axio-electric coupling $g_{Ae} = 10^{-11}$  and is proportional to ${g_{Ae}}^2$. The axion mass values correspond to the minimum energy value for each curve.}\label{fig:SolarAxionFlux}
\end{figure}

Our total flux is thus given by:
\begin{equation}
\frac{d\Phi}{dE}=\frac{d{\Phi}^{brem}}{dE}+\frac{d{\Phi}^{Comp}}{dE}
\end{equation}

where $\Phi^{brem}$ and $\Phi^{Comp}$ are the axion fluxes from Bremsstrahlung and Compton scattering, respectively. The emitted axion energy arises mainly from the energy of the electrons in the solar core, assumed to have a Maxwell-Boltzmann distribution, and has an average value of  5.1 keV for the Compton process and 1.6 keV for the Bremsstrahlung process, even for very low values of axion rest mass.

\subsection{Expected solar event rates from axio-electric effect}
\begin{figure}[htp]
\centering
\includegraphics[width=14cm] {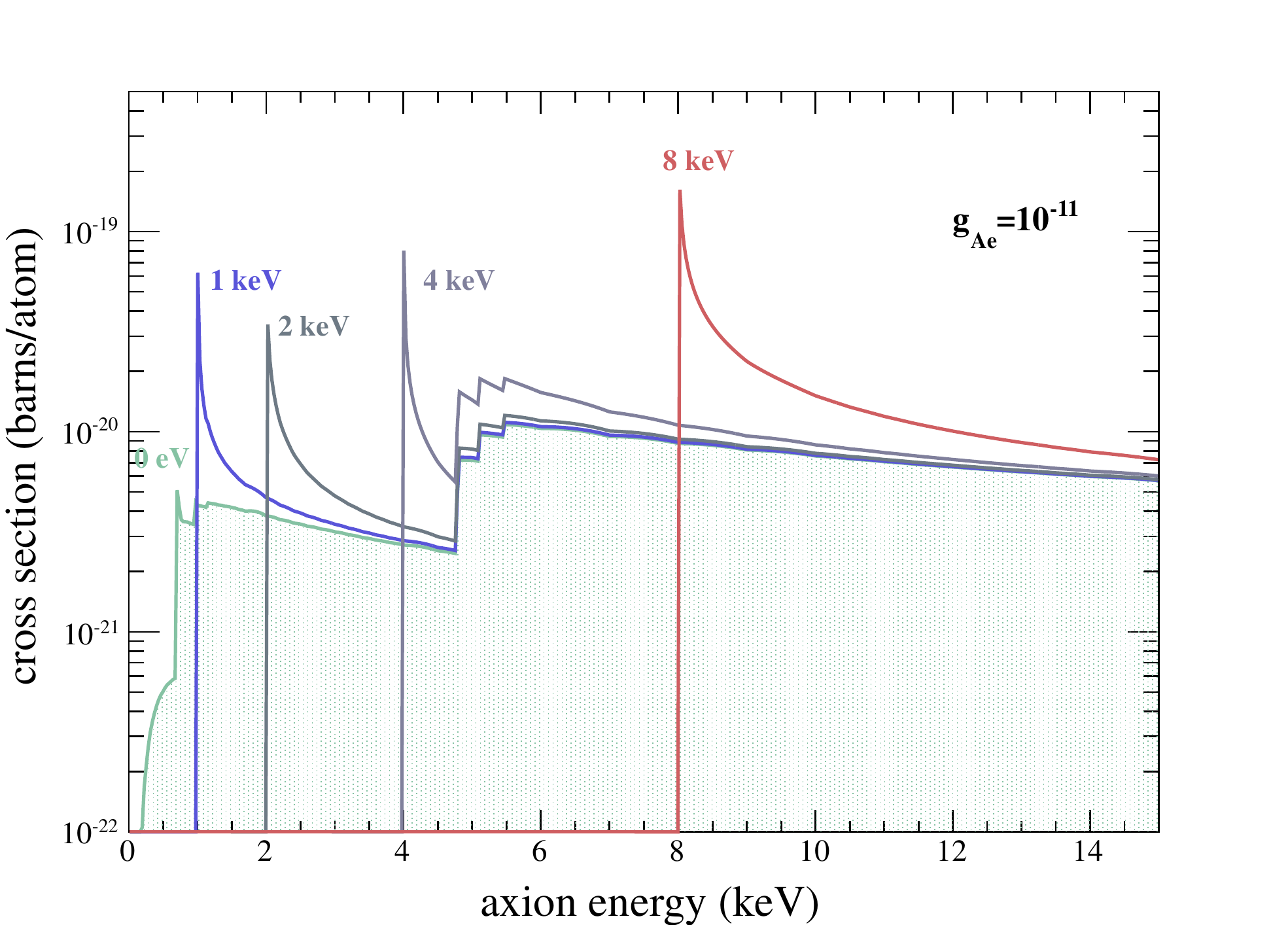}
\caption{Axio-electric cross sections for various axion masses as a function of axion total energy for a coupling  $g_{Ae}=1\times10^{-11}$. The green shaded region represents the case of massless axions.}\label{fig:SolarAxionCrossSection}
\end{figure}

 Event rates can be calculated from the convolution of the flux and axio-electric cross section. The axio-electric cross section Eq.~\ref{eqn:xs_A}, using the photoelectric cross section shown in Figure~\ref{fig:PhotoelectricCrossSection}, has the dependence on axion mass and axion total energy shown in Figure~\ref{fig:SolarAxionCrossSection}.

\begin{figure}[htp]
\centering
\includegraphics[width=14cm] {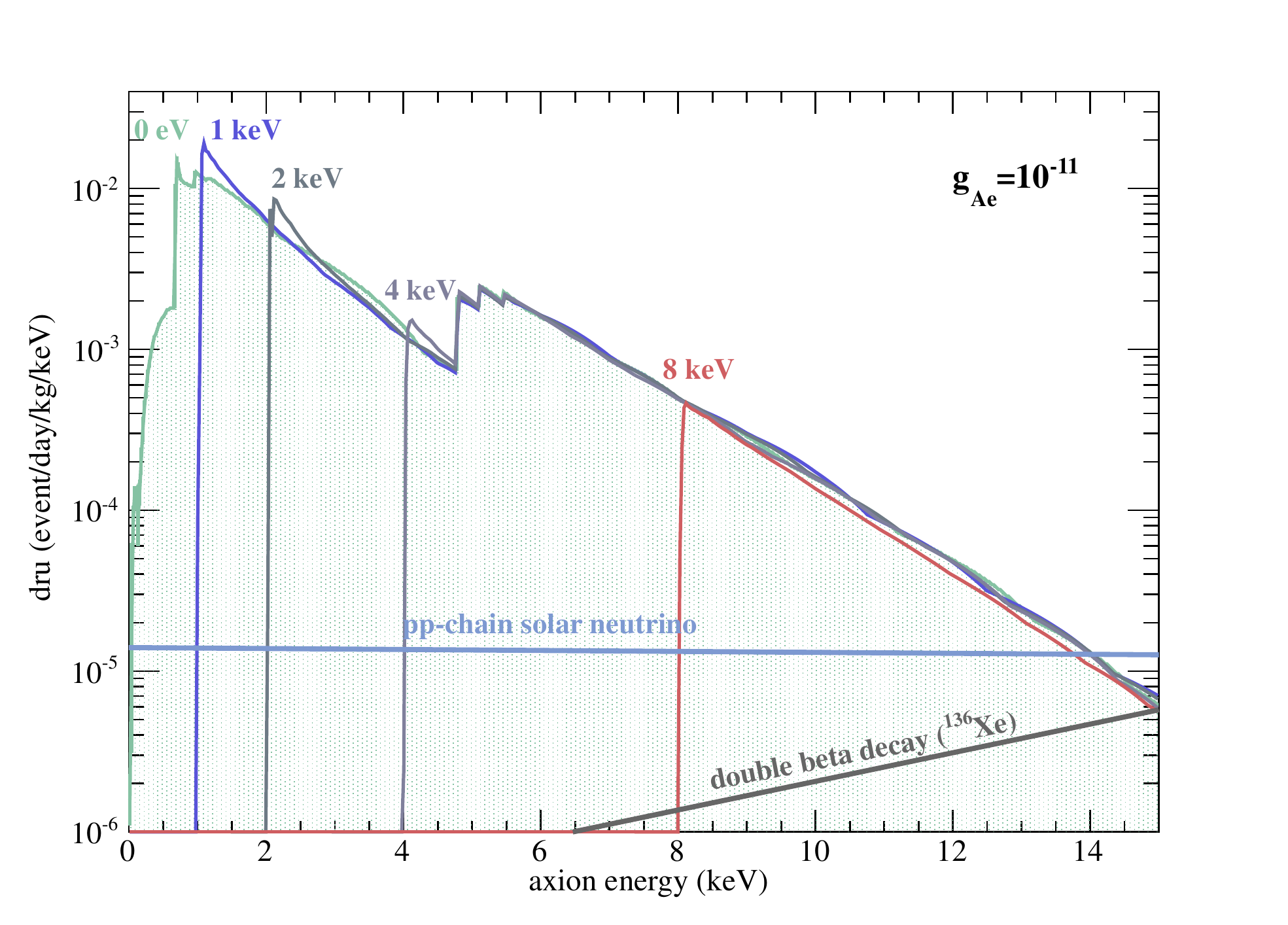}
\caption{Differential energy spectra (in units dru) in a Xe target for electrons emitted from axio-electric interactions for various axion masses as a function of axion total energy. The absolute scale corresponds to a coupling $g_{Ae} = 10^{-11}$,  and is proportional to $g_{Ae}^4$. Also shown here is the irreducible background due to pp-chain solar neutrinos~\cite{Bahcall:2005} and 2 $\nu$ double beta decay~\cite{Ackerman:2011}. The green shaded region represents the case of massless axions.}\label{fig:SolarAxionEventRate}
\end{figure}

Convolution of  the total flux versus energy from Figure~\ref{fig:SolarAxionFlux} with the energy-dependent cross sections in Figure~\ref{fig:SolarAxionCrossSection} then gives the expected differential energy spectra of electrons emitted by the axio-electric effect in a xenon target, shown in Figure~\ref{fig:SolarAxionEventRate}.  The spectrum is peaked close to the axion rest mass, from which it is evident that low axion masses are limited to the energy threshold of the detector. Note that the signal rate is proportional to ${g_{Ae}}^4$, since both the solar flux and the axio-electric signal are proportional to ${g_{Ae}}^2$.  For axion masses approaching zero, the spectrum is peaked around 1 keV - for this it is important to have an energy threshold $\sim$ 0.5 keV, a level close to that achieved in the ionization channel of from two-phase xenon dark matter detectors. Thus, from Figure~\ref{fig:SolarAxionEventRate}, the basic signal detection energy window will be the range 0.5 - 10 keV.  Also shown in Figure~\ref{fig:SolarAxionEventRate} are the irreducible backgrounds due to pp-chain solar neutrino and two-neutrino double beta decay at a level of 10$^{-5}$ dru.

\subsection{Determination of experimental limits on $g_{Ae}$ for solar axions}
 A signal would be present as a peak at an energy close to the axion mass, spread by the energy resolution. In the absence of a significant peak an experiment may set, at a chosen confidence level, a statistical limit based on the observed background in that energy region. The notation and procedure for detection or limit setting is similar to that described in Section~\ref{sec:ExperimentalLimits}. Again the energy resolution is defined by Eq.~\ref{eqn:resolution} and a resolution-defined window at each energy with the minimum upper energy set at 8 keV. The number of background events in this window replaces Eq.~\ref{eqn:nBG} for the purpose of defining the significance factor (signal-to-noise ratio) defined by Eq.~\ref{eqn:significance}.

The total axio-electric event rates, corresponding to each axion mass in the solar spectra of Figure~\ref{fig:SolarAxionFlux}, are obtained by numerical integration of the products of flux and cross section calculated for Figure~\ref{fig:SolarAxionEventRate} for the specific value $g_{Ae} = 10^{-11}$, the absolute value being proportional to $g_{Ae}^4$.  The latter can then be adjusted so that the events under the peak match the required confidence level $r_{sn}$, and the process repeated to provide a limit curve of $g_{Ae}$ versus axion mass. 

  Because of the required numerical integration over flux and photoelectric cross section $\sigma_{pe}$  to obtain a total signal event rate $S_{tot}$ , there is no analytical equivalent of Eq.~\ref{eqn:DMUpperLimit}, which showed the dependence on all input parameters in the case of dark matter axions. The nearest equivalent is
  \begin{equation}
\label{eqn:SolarUpperLimit}
 g_{Ae}(90\% C.L.)=1.4\times10^{-11}\left(\frac{r_{sn}}{S_{tot}}\right)^{\frac{1}{4}}\left[\frac{b\times(dN/dE)}{WT}\right]^{\frac{1}{8}}
  \end{equation}
 which shows the greater difficulty in this case of limit improvements via target mass, live time, low background, and photoelectron yield, all of which improve $g_{Ae}$  only as the eighth root (because of the dependence of solar rates on $g_{Ae}^4$ compared with $g_{Ae}^2$ for dark matter rates).
  
\section{Case studies:  application to G1, G2, and G3 experiments}
\label{sec:casestudies}
We consider four categories of detectors here, as shown in Table~\ref{table:InputParams}: G1 (Generation 1), G1.5, G2 and G3.
G1 is represented by the experiment with the best current WIMP results for intermediate mass WIMPs, XENON100. The G1.5 category consists of 100 kg-scale near-future experiments such as XMASS~\cite{Suzuki:2012} and LUX~\cite{Gibson:2012}. G2 is represented by a one ton scale detector such as XENON1T~\cite{XENON1T:2012} and G3 by a 10 ton scale detector such as XAX~\cite{Arisaka:2012_2D,Arisaka:2009}. For G1, the energy resolution at 1 keV of $b=0.6$ is derived from a simplified extrapolation of the observed resolution of XENON100, stated in~\cite{XENON1T:2012_1}:  $\sigma/E  = 1.9\%/\sqrt{E \mathrm{[MeV]}}$. For the single-phase G1.5 (XMASS), the resolution parameter $b=0.4$ is adopted, based on extrapolation of the XMASS observed resolution at 122 keV~\cite{Suzuki:2012}, and under the simplified and optimistic assumption of comparable scintillation yield at low energy.  For all of the future dual phase TPC experiments (G1.5, G2 and G3)  we also assume $b=0.4$, based on extrapolation of the observed energy resolution achieved in two-phase experiments and exploiting the superior resolution of the ionization channel~\cite{XENON1T:2012_1,Akimov:2012,Lebedenko:2009,Solovov:2011}.  

The electromagnetic background level is expected to be reduced substantially from generation to generation. We take 10 mdru for G1 as a conservative value from XENON100~\cite{Melgarejo:2012}. For G1.5, we assume 0.5 mdru as a conservative expectation from XMASS/LUX. For G2, it is assumed that the only background is in the form of irreducible neutrino interactions, stemming from pp-chain solar neutrinos and two neutrino double beta decay of $^{136}$Xe, which is shown in Figure~\ref{fig:SolarAxionEventRate} and can be parameterized as $dN/dE$ = $(1.4+0.07E)\times10^{-5}$ dru. For G3, we assume that $^{136}$Xe is depleted, and thus that only pp-chain solar neutrino contribute to the background.

As for the effective exposure (kg$\cdot$days) G1, $T$=1/2 year is obtained from 225 days of the latest XENON100 live time (together with $W$= 34 kg)~\cite{XENON100:2012} with an assumed 80\% efficiency. Otherwise, we simply assume $T$= 1 year for G1.5 and $T$= 2 years for G2 and G3.

  \begin{table}
  \caption{Experiment Input Parameters}
  \centering
\begin{tabular}{l l c c c c c}
\hline
&  & Weight & Live Time& Resolution & Energy Threshold & Background \\
 &  & $W$ & $T$& $b$& & $dN/dE$  \\  
 &  & (kg) & (years)& ($\sqrt{\mathrm{keV}}$) & (keV) & (dru) \\[0.5ex]
\hline\hline
G1 & XENON100 & 34 & $\frac{1}{2}$ & 0.6 & 2 & 0.01 \\
G1.5 &LUX/XMASS & 100 & 1 & 0.4& 1& $5\times10^{-4}$ \\
G2 & XENON1T & 1000 & 2 &0.4 & 1 &(1.4+.07E)$\times10^{-5}$$^{*}$\\
G3 &XAX & 10000 & 2 & 0.4& 1 &1.4$\times10^{-5}$$^{\dagger}$\\
\hline
\end{tabular}

\vspace{3mm}
 {\small $^{*}$pp $\nu$ chain+ 2$\nu$DBD of $^{136}$Xe\hspace{9mm}  $^{\dagger}$ pp $\nu$ chain only,$^{136}$Xe depleted}
\label{table:InputParams}
\end{table}

 Note that these assumptions for energy resolution, background rates, exposure and fiducial masses are likely overly simplistic, and may not accurately represent operational conditions. They are intended as conservative values to illustrate the potential sensitivity for the current and future experiments at 0.1-10 ton scale, although necessarily depend on various detector dependent factors not treated here.

\begin{figure}[htp]
  \centering
  \includegraphics[width=12cm] {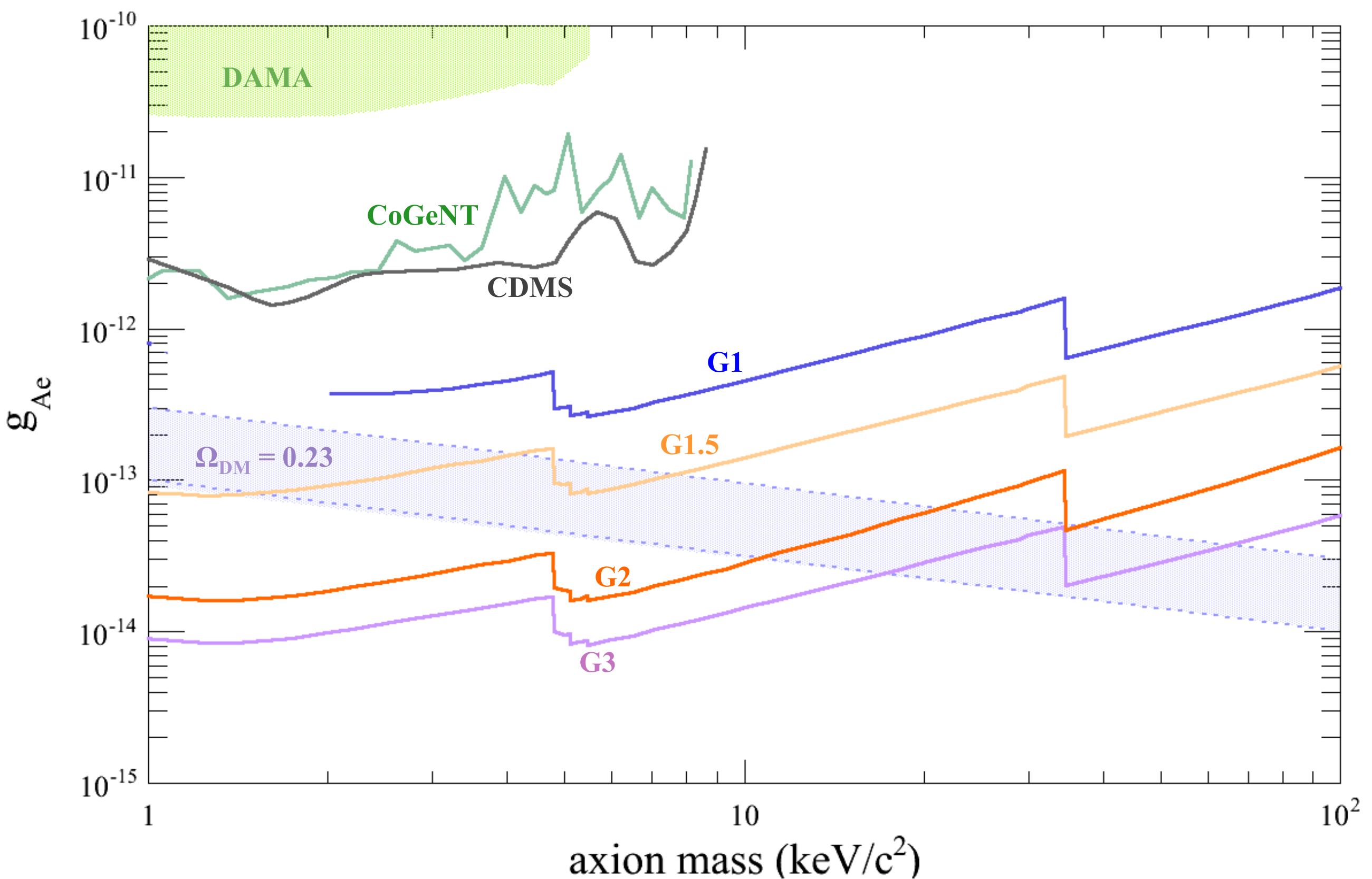}
  \caption{90\% CL limits on axion-electron coupling constant versus axion mass, based on reported G1 backgrounds together with the predicted sensitivity curves from expected backgrounds in G1.5, G2 and G3 detectors. The shaded band shows the predicted band for $g_{Ae}$ for ALPs under the assumption $\Omega_{DM}=0.23$. CDMS, CoGeNT, and DAMA data obtained from~\cite{Ahmed:2009,Aalseth:2008}.}\label{fig:90CLcomparison.pdf}
  \end{figure}

The resulting $90\%$ CL limits are shown in Figure 6 for $g_{Ae}$ versus axion mass for the different experiments listed in Table~\ref{table:InputParams}. Also shown in this figure are the results from CDMS, CoGeNT, and DAMA, obtained from~\cite{Ahmed:2009,Aalseth:2008}. As noted earlier, these limits apply to ALPs which have the potential to be the dark matter of the Universe. Indeed, in the the Figure we show the predicted band for $g_{Ae}$ in the case of ALPs giving $\Omega_{DM} = 0.23$. The shaded band in Figure~\ref{fig:90CLcomparison.pdf} uses the estimate for the ALPs density given in Eq. (16) of the  Pospelov et al paper~\cite{Pospelov:2008}, with the additional assumption that the ALPs couple to fermions universally with $L_{aff} = \frac{\partial^{\mu} a}{f} \bar{f}\gamma_{\mu}\gamma_5 f$, with $g_{Ae} = 2m_e/f$ .  As the figure shows, G1 is close to entering into this predicted band and G2-G3 should be able to test this theoretical prediction if the axion mass is less than 10 keV/c$^2$.

\begin{figure}[htp]
  \centering
  \includegraphics[width=14cm] {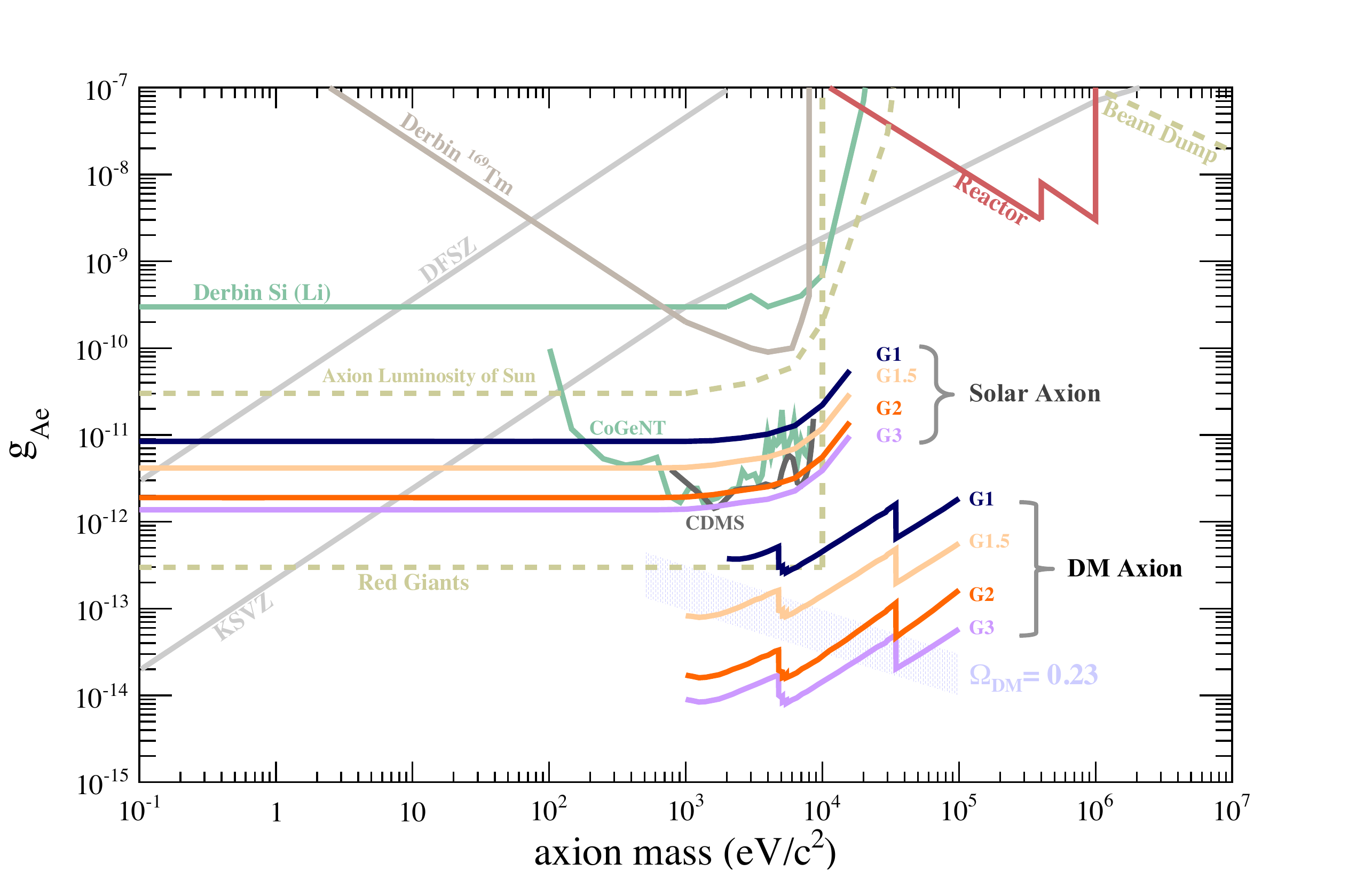}
  \caption{Summary of  limits on axion-electron coupling including limits from astrophysical sources and from other terrestrial experiments. The current and future liquid xenon experiment backgrounds can set the world best limit for a wide range of axion masses from massless up to 100 keV/c$^2$. The limits shown include astrophysical bounds from solar neutrino flux~\cite{Gondolo:2009} and red giants~\cite{Raffelt:2006}, dedicated axion experiments by Derbins using  $^{169}$Tm~\cite{Derbin:2011} and Si(Li)~\cite{Derbin:2012}, reactor/beam dump experiments~\cite{Konaka:1986,Chang:2007,Bellini:2008}, and theoretical bounds~\cite{Kaplan:1980}.}\label{fig:gaelimits3.pdf}
  \end{figure}

In Figure~\ref{fig:gaelimits3.pdf}, the relation between $g_{Ae}$ and axion mass is extended to a much wider region than in Figure~\ref{fig:90CLcomparison.pdf} to show current and projected limits from both solar and dark matter axions by the G1-G3 liquid xenon experiments with predictions of the DFSZ and KSVZ  invisible axion models~\cite{Kaplan:1980} where $m_A = 6$  $\rm{eV}\times(10^{6}/\textit{f} [GeV])$. Clearly, the G2/G3 experiments will be able to set a stringent upper bound on the axion mass for the DFSZ and KSVZ models for the mass ranges $m_A < 0.1$ eV/c$^2$and $<$ 10 eV/c$^2$, respectively. We note that the range of values of $g_{Ae}$ that can be probed by the G1-G3 experiments corresponds to a symmetry-breaking scale $ 10^8 $ $\rm{GeV}<$ \textit{f} $ <10^9$ $\rm{GeV}$ which is one to two orders of magnitude more sensitive than what is achievable at CAST~\cite{Andriomonje:2007} for low mass axions.
 
Figure~\ref{fig:gaelimits3.pdf} also shows astrophysical luminosity limits from the solar neutrino flux of the sun~\cite{Gondolo:2009} and red giants~\cite{Raffelt:2006}, experiments by Derbin et al.~\cite{Derbin:2011,Derbin:2012}, reactor experiments~\cite{Bellini:2008,XENON1T:2012}, beam dump experiments~\cite{Konaka:1986}, as well as CoGeNT and CDMS~\cite{Ahmed:2009}. This figure demonstrates that liquid xenon detectors are potentially capable of setting world-leading limits on the axio-electric coupling for the entire range of axion masses from zero to 100 keV/c$^2$. 

\begin{figure}[htp]
  \centering
  \includegraphics[width=12cm] {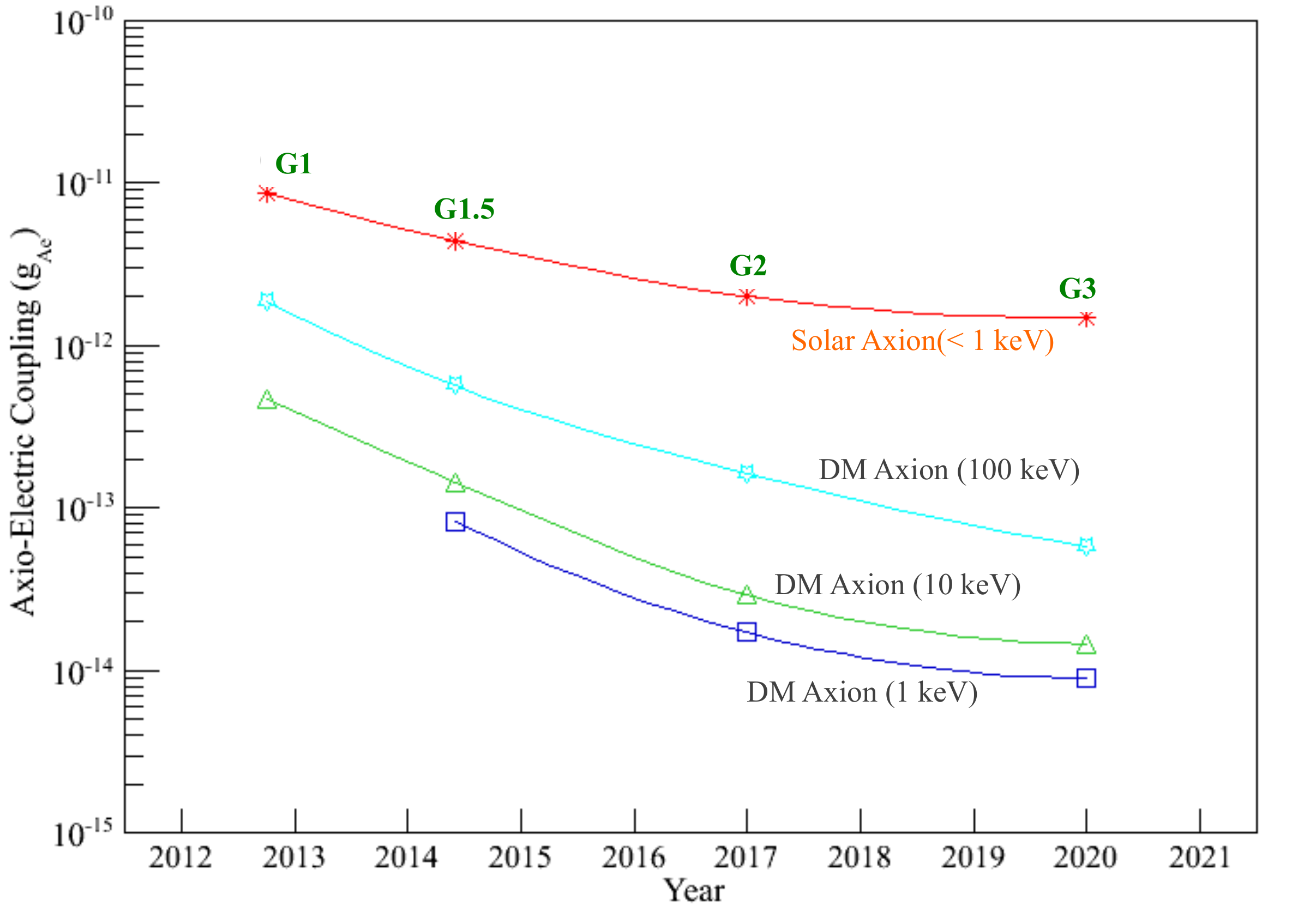}
  \caption{Potential improvements with time of the axion-electron coupling $g_{Ae}$ from current and future liquid Xenon experiments.}\label{fig:gaelimits1.pdf}
  \end{figure}

Figure~\ref{fig:gaelimits1.pdf} shows potential improvements in limits as a function of calendar years.  Lastly, a summary of experimental parameters, rates, and limits is given in Table~\ref{table:BigTable} (the entry for vector boson dark matter is calculated in Section 5).

\newgeometry{bottom =2.50cm, right= 3.00cm, left=3.80cm, top=3.0cm}
\begin{scriptsize}
\begin{landscape}\begin{table}\caption{Input Parameters and Obtained Limits}\end{table}
{\small  
\begin{longtable}{lp{7.75cm}rcccccc}
\hline
& & & \multicolumn{4}{c}{Experiment}\\
\hline
&&&G1&G1.5&G2&G3\\
 Symbol & & Unit & \textsc{XENON}100 & LUX/XMASS & \textsc{XENON}1T & XAX \\\hline\hline
\endhead
& \textit{Year} & & 2012.9  & 2014.5 & 2017 & 2020\\
W & \textit{Fiducial Mass} & kg & 34 & 100 & 1000 & 10000\\
T & \textit{Effective Run Time} & days & 180 & 365 & 730 & 730\\
b & \textit{Energy Resolution at 1 keV} & & 0.6 & 0.4 & 0.4 & 0.4\\
& \textit{Energy Threshold} & keV & 2 & 1 & 1 & 1\\\hline
\multicolumn{9}{l}{Galactic Axion (Mass = 1 keV, $\sigma_{pe}=2.05\times10^{6}$ barns/atom)}\\\hline
S & \textit{No. of Signals (for $g_{Ae} = 1\times10^{-12}$}) & events & - & 2734 & 58590 & 546848\\
dN/dE & \textit{Background rate} & /kg/day/keV & - & $5.00\times10^{-4}$ & $1.50\times10^{-5}$ & $1.40\times10^{-5}$\\
N & \textit{No. of Backgrounds} & events & - & 127 & 88 & 715\\
& \textit{90\% CL Limit on $g_{Ae}$} & & - & $8.29\times10^{-14}$ & $1.72\times10^{-14}$ & $9.04\times10^{-15}$\\
& \textit{90\% CL Limit on $\alpha^{'}/\alpha$ for Vector DM} & & - & $2.06\times10^{-31}$ & $8.89\times10^{-33}$ & $2.44\times10^{-33}$\\
\hline
\multicolumn{9}{l}{Galactic Axion (Mass = 10 keV, $\sigma_{pe}=3.62\times10^{4}$ barns/atom)}\\\hline
S & \textit{No. of Signals (for $g_{Ae} = 1\times10^{-12}$}) & events & 187 & 894 & 19148 & 178711\\
dN/dE & \textit{Background rate} & /kg/day/keV & $1.00\times 10^{-2}$ & $5.00\times10^{-4}$ & $2.10\times10^{-5}$ & $1.40\times10^{-5}$\\
N & \textit{No. of Backgrounds} & events & 585 & 117 & 80 & 522\\
& \textit{90\% CL Limit on $g_{Ae}$} & & $4.64\times10^{-13}$ & $1.44\times10^{-13}$ & $2.90\times10^{-14}$ & $1.46\times10^{-14}$\\
& \textit{90\% CL Limit on $\alpha^{'}/\alpha$ for Vector DM} & & $6.75\times10^{-28}$ & $6.54\times10^{-29}$ & $2.66\times10^{-30}$ & $6.72\times10^{-31}$\\
\hline
\multicolumn{9}{l}{Galactic Axion (Mass = 100 keV, $\sigma_{pe}=3.91\times10^{2}$ barns/atom)}\\\hline
S & \textit{No. of Signals (for $g_{Ae} = 1\times10^{-12}$}) & events & 21 & 99 & 2133 & 19913\\
dN/dE & \textit{Background rate} & /kg/day/keV & $1.00\times 10^{-2}$ & $5.00\times10^{-4}$ & $8.70\times10^{-5}$ & $1.40\times10^{-5}$\\
N & \textit{No. of Backgrounds} & events & 1833 & 365 & 1015 & 1634\\
& \textit{90\% CL Limit on $g_{Ae}$} & & $1.85\times10^{-12}$ & $5.65\times10^{-13}$ & $1.63\times10^{-13}$ & $5.82\times10^{-14}$\\
& \textit{90\% CL Limit on $\alpha^{'}/\alpha$ for Vector DM} & & $1.10\times10^{-24}$ & $1.03\times10^{-25}$ & $8.63\times10^{-27}$ & $1.08\times10^{-27}$\\
\hline
\multicolumn{9}{l}{Solar Axion (Mass $<$1 keV, $g_{Ae}=1\times10^{-11}$, $S_{r}$=0.02 event/kg/day)}\\\hline
S & \textit{No. of Signals (for $g_{Ae} = 1\times10^{-12}$}) & events & 66 & 487 & 9989 & 98890\\
dN/dE & \textit{Background rate} & /kg/day/keV & $1.00\times 10^{-2}$ & $5.00\times10^{-4}$ & $1.40\times10^{-5}$ & $1.40\times10^{-5}$\\
N & \textit{No. of Backgrounds} & events & 458 & 128 & 88 & 883\\
& \textit{90\% CL Limit on $g_{Ae}$} & & $8.58\times10^{-12}$ & $4.43\times10^{-12}$ & $2.00\times10^{-12}$ & $1.49\times10^{-12}$\\
\label{table:BigTable}
\end{longtable}}
\end{landscape}
\end{scriptsize}
\restoregeometry
\normalsize

  \section{KeV-scale Vector Boson Dark Matter}
\subsection{Bosonic Super-WIMP models}
 Pospelov et al.~\cite{Pospelov:2008} discuss the possibility of new types of vector bosons in the keV/c$^2$ mass range, which couple to photons or other light particles sufficiently weakly to ensure survival times exceeding the present lifetime of the Universe, and thus are sufficiently stable to constitute the dark matter. These particles, like the ALPs, would also be detectable via the axio-electric effect as a result of a very small dimensionless coupling $\alpha '$ to electrons, analogous to the standard coupling $\alpha(=e^2/4{\pi}hc)$ of photons to electrons. Absence of an observable signal would thus enable a limit to be set on the ratio $\alpha '/\alpha$

\subsection{Procedure for calculation of limits for bosonic super-WIMPs}
  From Pospelov et al. Eq. (40), the signal rate $S_R$ (events/kg/day) for a dark matter vector boson of mass $m_V$ (keV/c$^2$) would be
  \begin{equation}
  S_R=\left(\frac{4\times10^{23}}{A}\right)\left(\frac{\sigma_{pe}}{m_V}\right)\left(\frac{\alpha '}{\alpha}\right)
  \end{equation}
  where the photoelectric cross section $\sigma_{pe}$ is in barns/atom.
  Thus the total signal events in a Xe target would be
  \begin{equation}
\label{eqn:nBG_VectorDM}
  N_{tot}=3\times10^{21}WT\left(\frac{\sigma_{pe}}{m_V}\right)\left(\frac{\alpha '}{\alpha}\right)
  \end{equation}
   As before, the signal energy window will be governed by the energy resolution $\sigma_{E_e}$ from Eq.~\ref{eqn:resolution} and extend $\pm 2\sigma_{E}$ from the boson mass $m_V$ (keV/c$^2$), for which the number of background events is given by Eq.~\ref{eqn:nBG}. Comparing this with Eq.~\ref{eqn:nBG_VectorDM} we determine the limit on $(\alpha '/\alpha)$ for a given significance level Eq.~\ref{eqn:significance}:  \begin{equation}
\label{eqn:VectorDMUpperLimit}
  \frac{\alpha'}{\alpha}=1.0\times10^{-23}{m_V}^{\frac{5}{4}}\left(\frac{r_{sn}}{\sigma_{pe}}\right)\left[\frac{b\times(dN/dE)}{WT}\right]^{\frac{1}{2}}
  \end{equation}
  (with $r_{sn}$ = 1.7 for 90\% confidence).
  
\subsection{Case studies for G1, G2, and G3 experiments}
\begin{figure}[htp]
  \centering
  \includegraphics[width=14cm] {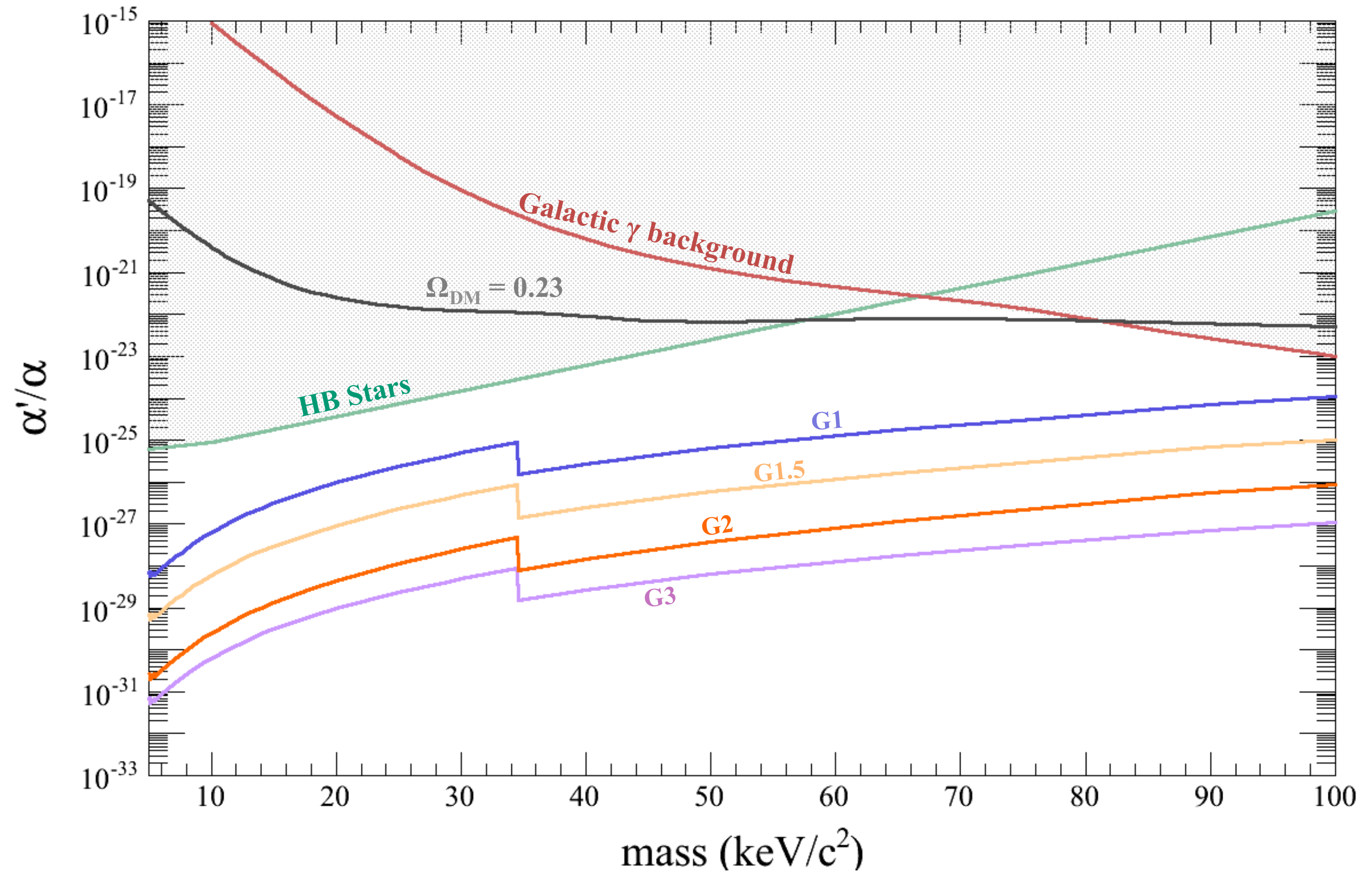}
  \caption{Limits on vector boson super-WIMP coupling compared with astrophysical limits. Other limits include those from HB stars~\cite{Aalseth:2008}, theoretical dark matter abundance, and galactic $\gamma$-background~\cite{Pospelov:2008}.}\label{fig:vectorcouplinglimits1.pdf}
  \end{figure}

\begin{figure}[htp]
  \centering
  \includegraphics[width=12cm] {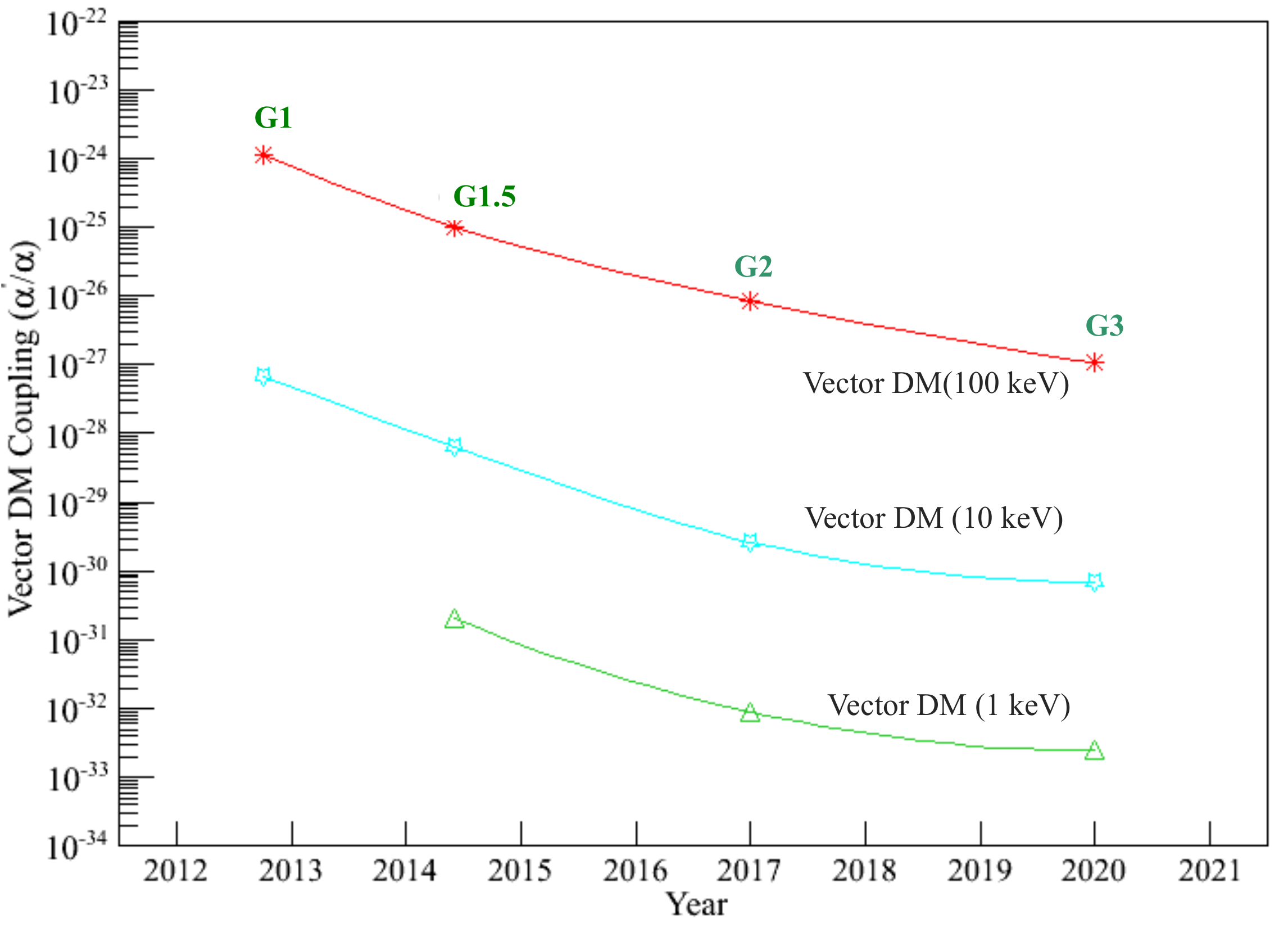}
  \caption{Projected improvements in sensitivity of the vector DM coupling $\alpha'/\alpha$ from current and future liquid Xenon experiments.}\label{fig:VectorDMExperiments.pdf}
  \end{figure}

  Using the above procedure, we show in Figure~\ref{fig:vectorcouplinglimits1.pdf} the limits on dark matter vector boson couplings resulting from the current and future background levels in G1-G3 experiments. These are compared with several astrophysical and cosmological limits (from~\cite{Pospelov:2008,Aalseth:2008}). As shown here, successful G1 experiments (such as XENON100) can already exclude the possibility of these vector bosons being the dominant form of dark matter of the Universe, since the G1 limit exceeds the expected value of $\alpha '/\alpha$ giving $\Omega_{DM}$ = 0.23 by a large margin. Figure~\ref{fig:VectorDMExperiments.pdf} summarizes potential improvements in $(\alpha '/\alpha)$ with calendar years.
  
  \section{Discussion and Conclusions}
\subsection{Discussion}
There are four parameters which characterize the detectors: $b$, $dN/dE$, $T$, and $W$ (see Table~\ref{table:nonlin}).

Using these quantities and Eqs.~\ref{eqn:DMUpperLimit},~\ref{eqn:SolarUpperLimit}, and~\ref{eqn:VectorDMUpperLimit}, the sensitivity to $g_{Ae}$ or ($\alpha '/\alpha$) can be expressed as the following:
 \begin{equation}
\label{eqn:GeneralUpperLimit}
  g_{Ae}\hspace{1mm}(\mathrm{or}\hspace{1mm} \alpha '/\alpha)\propto\left(\frac{b\times(dN/dE)}{WT}\right)^k
  \end{equation}
  
where 

\hspace{6mm}$k$=1/4 for $g_{Ae}$ in the case of DM (Galactic) axions

\hspace{6mm}$k$=1/8 for $g_{Ae}$ in the case of solar axions

\hspace{6mm}$k$=1/2 for $\alpha '/\alpha$ in the case of DM Vector bosons

Note that $\alpha$ and $\alpha '$ are actually proportional to the square of the fundamental vertex coupling, and so are equivalent to ${g_{Ae}}^2$. If one instead set limits on the ``charge ratio'' $e'/e$, then $k$ would be 1/4, as in the case of $g_{Ae}$.

\vspace{2mm}

\normalsize Eq.~\ref{eqn:GeneralUpperLimit} is of value in providing a means of comparing experiments, and of estimating improvements possible with larger experiments. 

\subsection{Conclusions}
In recent years, liquid xenon detectors have shown remarkable success in advancing the sensitivity of experiments to WIMP-nucleon interactions through searches for nuclear recoils in extended exposures.  In increasing the mass scale, lowering energy thresholds, and lowering backgrounds through self-shielding and 3D position reconstruction, these instruments now have sufficiently low electromagnetic background to provide capability to address a number of dark matter candidates in addition to the favoured WIMP hypotheses, reviving and potentially realizing searches through analogues of the photoelectric effect to detect weakly interacting light bosons~\cite{Cheng:1987,Dimopoulos:1986,Avignone:1987}.  We have shown that one can achieve new limits on axion and axion-like particles, as well as other boson couplings, which are sufficient to exclude a number of theoretical models, and extend dark matter searches with these instruments beyond primary nuclear recoil WIMP searches.  Moreover, this can be achieved without new designs for detectors - those already operating, under construction, or planned for WIMP detection can, without modification, set limits on axio-electric signals during the same WIMP searches used to address nuclear recoil signatures.  We have provided examples based on the background data from the existing or proposed projects showing that one can achieve exclusion limits with these that are lower than either astrophysical limits or  those coming from other types of experiment.

\section*{Acknowledgements}
We thank V. Derbin and A. Derevianko for useful discussions. We also thank Shigetaka Moriyama from the XMASS collaboration for fruitful dialogue. This work was supported by DOE Grant DE-FG02-91ER40662.


\bibliographystyle{apsrev4-1}

\end{document}